# All-optical Image Identification with Programmable Matrix Transformation

Shikang Li[1], Baohua Ni[1], Xue Feng[1, *], Kaiyu Cui[1], Fang Liu[1], Wei Zhang[1], and Yidong Huang[1]

[1] *Department of Electronic Engineering, Tsinghua University, Beijing, China*
*\* x-feng@tsinghua.edu.cn*

**Abstract:** An optical neural network is proposed and demonstrated with programmable matrix transformation and nonlinear activation function of photodetection (square-law detection). Based on discrete phase-coherent spatial modes, the dimensionality of programmable optical matrix operations is 30~37, which is implemented by spatial light modulators. With this architecture, all-optical classification tasks of handwritten digits, objects and depth images are performed on the same platform with high accuracy. Due to the parallel nature of matrix multiplication, the processing speed of our proposed architecture is potentially as high as 7.4T~74T FLOPs per second (with 10~100GHz detector).

## 1. Introduction

Optical neural network (ONN) have attracted more interest since it is very promising to perform the machine learning algorithm [1]. Mapping the large-scale parallel processing and massive interconnections onto optical hardware could evidently reduce the time and energy consumption of data processing. In the neural network, both linear and non-linear operations are required. For the linear operations, the interconnections of artificial neurons can be mathematically represented with matrix-vector multiplications. Till now, optical matrix multiplication schemes have been proposed involving the silicon photonic platforms [2,3], free-space diffractive systems [4–6], time-domain spiking neurons [7], frequency-domain convolutional accelerators [8], etc. Among these approaches, the matrix multiplications may be programmable or not. Obviously, if the programmable arbitrary matrix multiplications are employed, wider range of tasks can be implemented with ONN scheme.

For the non-linear operations (activation function), the techniques of photonic nonlinear neurons include the optical/electrical/optical (O/E/O) approach [9,10] and all-optical neurons with carrier regeneration [7,11]. The all-optical nonlinear neurons would potentially be faster but less power-efficient since the optical nonlinearity is very weak [1]. Recently, the deep diffractive neural network ($D^2NN$) [4] have been reported with multi-layer linear network as well as only one nonlinear layer achieved by the square-law detection of photodetectors. According to experimental results, considerable accuracy for image identification tasks has been achieved. Such results of $D^2NN$ indicated that it is an efficient method to implement ONN with linear network and photodetection since the photodetectors are quite mature devices.

In our previous works [12,13], programmable arbitrary linear optical operations have been proposed and demonstrated on discrete phase-coherent spatial modes. Thus, in this work, we proposed and demonstrated a programmable ONN scheme for various image identification tasks. In our scheme, the images are encoded on optical spatial modes so that programmable arbitrary complex matrix operations can be performed with high dimensionality, while the nonlinear neurons are achieved also by the square-law detection of photodetectors. With this architecture, the processing speed could be very fast due to the parallel nature of matrix multiplication and is actually limited by the adopted photodetector. As discussed later, our proposed ONN could achieve the speed as high as 7.4T~74T FLOPs per second with 10~100GHz detector.

## 2. Architecture

An all-optical architecture is proposed with a fully connective linear transformation followed by nonlinear activation function. Here, the simplest case of ONN is constructed, where the input artificial neurons are connected to the output artificial neurons through a fully connective complex matrix transformation. Next, the nonlinear activation function is achieved by detector array aligned to each output neurons. Figure. 1 illustrates the conceptual scheme that performs a typical task of image identification. The test image is directly encoded on the transverse amplitude distribution of the incident coherent light beam, as the original input of the ONN. Next, the input state is obtained through spatial sampling of the image and denoted as input state $|\alpha\rangle$ under discrete coherent spatial (DCS) mode basis [12], which is a group of individual beam spots that arranged arbitrarily within the two-dimensional transverse plane according to the optical beam propagation. As shown in Fig.1, the DCS mode basis can be readily employed to encode the spatially sampled image. Moreover, as presented later, the dimensionality as well as arrangement can be properly optimized according to the task.

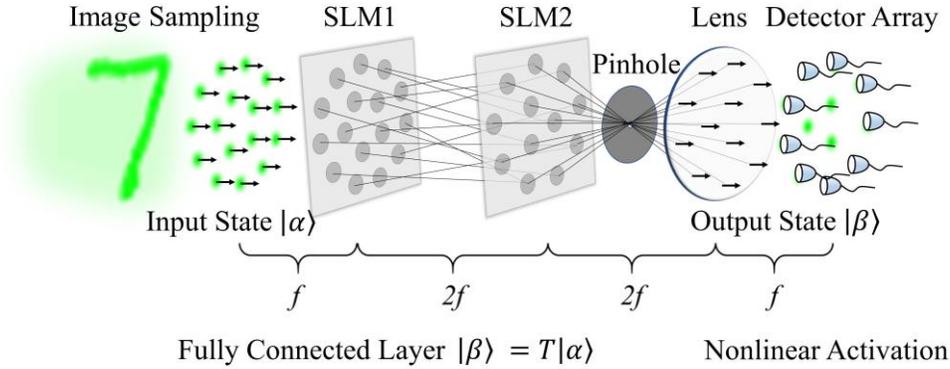

Fig.1. Conceptual scheme for all-optical image identification.

After the spatial sampling, our proposed ONN architecture operates in two steps:

$$\begin{aligned}linear : |\alpha\rangle \rightarrow T|\alpha\rangle = |\beta\rangle, \\ nonlinear : |\beta\rangle \rightarrow \||\beta\rangle\|^2\end{aligned} \quad (1)$$

Here, the nonlinear activation function is achieved by the photodetector array. It has been presented that the square-law detection of photodetectors could provide the required nonlinearity for ONN [4]. For the linear layer, the transformation $T$ between the input state $|\alpha\rangle$ and output state $|\beta\rangle$ is implemented with our previous works, in which a programmable complex matrix transformation scheme has been proposed with the aforementioned DCS modes and applied to both classical optical matrix transformation of 26 dimensionalities [13] and quantum projective measurements of 15 dimensionalities [12]. As shown in Fig. 1, this technique is based on meticulously designed phase-only modulation functions acting on wave front of each individual DCS modes. And only two spatial light modulators (SLMs) are needed to perform the one-to-all beam splitting and all-to-one beam recombining in parallel. The complex matrix elements are mapped to tunable splitting ratio and recombining ratio. There are two distinctive features in our design. First, the achievable matrices are not constrained by the unitary class like the integrated photonic scheme [2,14]. Thus, the non-unitary matrices and rectangular matrices, which are more promising and suitable to be employed in ONN applications, can be directly implemented without singular value decomposition (SVD). Second, the input vector $|\alpha\rangle$ is defined on DCS modes within the two-dimensional transvers plane. This assures exceptional compatibility to image processing tasks since the image signals could be processed immediately with this all-optical architecture, while there is no requirement of extra analog to digital convertors (ADC) and optical modulators in signal preparation.

It should be mentioned that various image identification tasks can be flexibly performed on the same platform as shown in Fig.1 due to the programmable matrix transformation. In this work, the image identification tasks from the MNIST (Modified National Institute of Standards and Technology) [15] handwritten digit database and MNIST fashion database are utilized to experimentally evaluate our proposed all-optical architecture.

The number of input neurons equals to the number of spatial sampling points of the test image, while there are 10 output neurons according to the number of labels of MNIST digit and MNIST fashion tasks. The predicted label of the test image is determined by the most intense output neuron. Thus, the accuracy of classification can be quantified. The transformation matrix is pre-trained before implementing with SLMs. During training progress, the loss function is defined as

$$F_L = \max(0, a - p) \qquad (2)$$

Here $p$ is the percentage of the intensity in the correct output detector region and $a$ is a scalar threshold according to the task. Typically, we adopted $a = 0.7$ for MNIST digit and $a = 0.9$ for MNIST fashion. The total cost function is calculated by summing loss functions of each individual image. The original dimensionality of MNIST digit or MNIST fashion image is $28 \times 28$. First, the down-sampling is made to decrease the number of input neurons as well as the matrix dimension. Next, the training image data are fed in batches. As shown in Fig. 2(a), the training takes several rounds. The optimization of the matrix is done with MATLAB 'fmincon' toolbox with the SQP (sequential quadratic programming) algorithm. Specifically, each round contains two epochs and both of them uses all the training data. In the first epoch, there are 60 batches while the batch size is 1000 and iteration limit of SQP algorithm is 5000. In the second epoch, the batch size is increased to 2000 and the iteration limit is changed to 3000. The batch size and iteration limit during training are varied to learn rapidly in the first epoch and then stabilize the model in the second one. For the testing process, all 10000 test data are fed. The accuracy as a function of down-sampling level (or number of input neurons) for both tasks are plotted in Fig. 2(b) and (d). The input dimensionalities of both networks are determined according to these numerical calculations. For MNIST digit and fashion dataset, the input dimensionalities are 30 and 37, respectively, while the output dimensionalities are the same as 10. The spatial distribution of sampling points of these two configurations are shown in Fig. 2(c) and (f), respectively. The reason for choosing the circularly symmetric alignment of spatial sampling is to fit the circle apertures of optical elements. With respect to the well-distributed sampling points in Fig. 2(c) and (f), the training and testing results are shown in Fig. 2(d). It can be found that one round is enough and there are no obvious accuracy variations for more training rounds. It also indicates that the input dimensionalities of both networks are proper. For both tasks the difference between the training accuracy of 60000 data and testing accuracy of 10000 data is less than 1%. Without much loss of accuracy, the networks after one training round as shown in Fig.2(d) are implemented in experiment. The corresponding image identification accuracy in numerical blind testing is 88.76% for MNIST digit and 78.44% for MNIST fashion over 10000 test data.

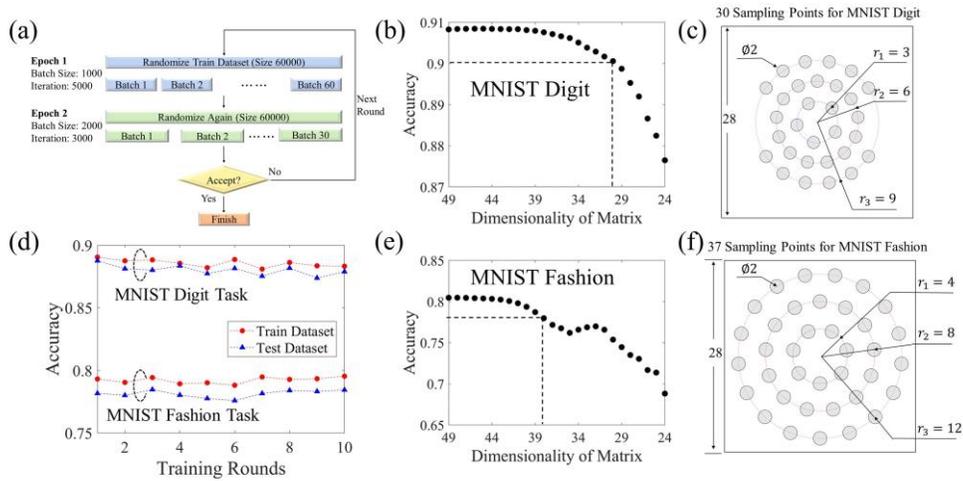

Fig.2. (a) The flow chart of training process. (b) and (e) Accuracy as a function of network dimensionality for MNIST digit task and MNIST fashion task, respectively. (c) and (f) The employed distribution of 30 sampling points for MNIST digit task and 37 sampling points for MNIST fashion task, respectively. (d) Accuracy of training and blind testing for both tasks as a function of training rounds.

## 3. Experiments and Results

The experimental setup is shown in Fig. 3(a). A distributed Bragg reflector (DBR) laser operating at 808 nm (Thorlabs, DBR808PN) with 1 MHz linewidth is injected to the free space optical system through a fiber collimator. The polarization control consists of a half wave plate (HWP) and a linear polarizer to meet the requirement of SLMs. In Fig. 3(a), SLM0 serves for input image generation, while SLM1 and SLM2 performs the optical matrix transformation with the help of a pinhole to construct the fully connective linear layer of ONN. This experimental setup is the same as the architecture shown in Fig. 1, except for the SLMs work in reflective mode. The phase masks settled on SLM1 for MNIST digit task and MNIST fashion task are shown in Fig. 3(b) and (c), respectively. There are 30 and 37 active regions on SLM1 in Fig. 3(b) and (c) according to the distribution of spatial sampling points shown in Fig. 2(c) and (f), respectively. Each active region performs the one-to-ten beam splitting of programmable complex splitting ratio in parallel, while these different splitting ratios could fulfill the required $30 \times 10$ (or $37 \times 10$) elements of network weights. Fig. 3(c) shows the zoomed picture of a typical active region, where gray scale is proportional to phase modulation. Besides the active regions, the other regions on SLM1 is set to be irreflexive with the checkerboard method, in which the adjacent pixels are settled as zero and π. Thus, spatial sampling of the test image as well as beam splitting are done at the same time by SLM1. The SLM employed in this work (Holoeye GAEA-2 series) has a physical size of $8 \times 14\ mm^2$ and spatial resolution of $2160 \times 3840$. The MNIST fashion images with intensity coding are projected onto the SLM1 plane with size of $8 \times 8\ mm^2$ according to the shorter edge of SLM. Since the MNIST digit images are statistically more centralized, for MNIST digit task the generated images are enlarged by 30%. A sketch of the phase mask on SLM2 is shown in Fig. 3(d). For both tasks, there are ten active regions on SLM2 and all of them perform 30-to-one (or 37-to-one) beam recombining in parallel. These ten active regions on SLM2 are marked as zero to nine according to the ten labels of image identification tasks. For MNIST digit task, the mapping is simply to the same digits. For MNIST fashion task the mapping is t-shirts, trousers, pullovers, dresses, coats, sandals, shirts, sneakers, bags, and ankle boots for label zero to nine. The matrix output is measured by a charge coupled device (CCD) camera and the intensity within 10 detector regions are recorded to determine the predicted label. The nonlinear activation function in Eq. (1) is achieved by intensity measurement and applied to the ten output neurons. In Fig. 3 (e) and (f), the measured output intensity distribution of digit four and fashion trousers are shown. Ten detector regions are marked by dashed read circles. The test images of digit and fashion inserted in Fig. 3(e) and (f) are recorded by another CCD camera with the help of a flip mirror (not shown in Fig. 3(a)) placed between SLM0 and SLM1.

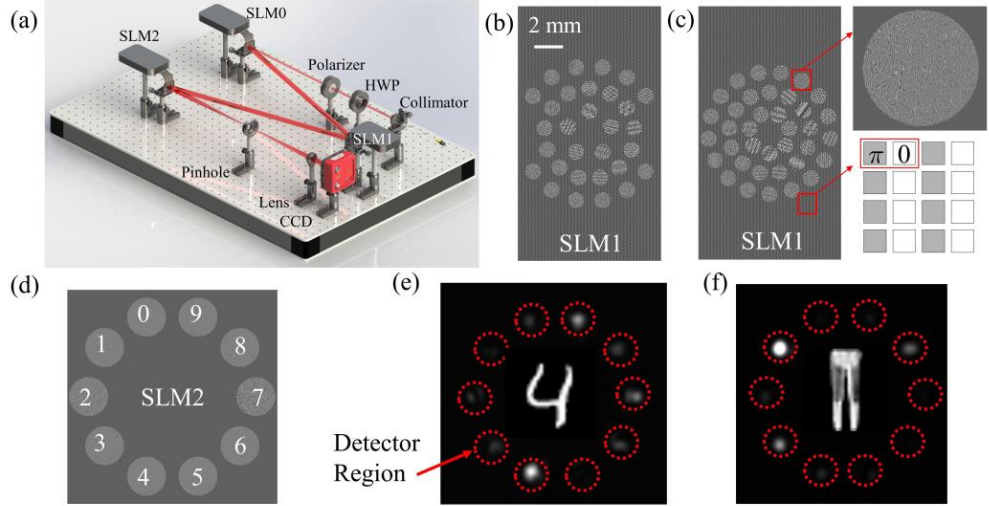

Fig.3. (a) Sketch of experimental setup. (b) and (c) Phase modulation functions settled on SLM1 for MNIST digit and MNIST fashion tasks, respectively. (d) Phase modulation functions settled on SLM2. (e) Measured intensity within ten detector regions for digit identification. (f) Measured intensity within ten detector regions for fashion identification.

The performance of all-optical image identification has been evaluated by experimentally classifying the first 200 out of 10000 testing images for both MNIST digit database and MNIST fashion database. The results about MNIST digit task are summarized in Fig. 4. The complex amplitude of the optimized linear network weight is shown in Fig. 4(a), where the height and color of histograms indicate absolute value and phase, respectively. Such $30 \times 10$ network could achieve an accuracy of 88.76% as mentioned before. The confusion matrix and energy distribution percentage according to numerical calculation are shown in Fig. 4(b) and (c). In Fig. 4(d), output intensity distribution of the 1~50 from 200 experiments are shown, where filled bars and empty bars indicate measured intensity and theoretical intensity calculated from target network, respectively. The true labels are also marked in Fig. 4(d) while correct and incorrect classification are noted in blue and red, respectively. To quantify the implementation accuracy of neural network, the statistical fidelity is calculated according to [16]:

$$F_s(P_{\exp}, P) = \sum \sqrt{P_{\exp} \cdot P} \qquad (3)$$

The label $P_{exp}$ and $P$ denote the measured intensity distribution and theoretical intensity distribution, respectively. The fidelity of this $30 \times 10$ network implementation is $0.949 \pm 0.026$ over 200 tests. The experimental confusion matrix and energy distribution percentage are shown in Fig. 4(e) and (f). Since the test images are not pre-selected, each true label does not have equal occurrence. However, it can be seen that the experimental results are in similar tendency with the numerical calculation. Totally 170 correct classification are observed among 200 tests.

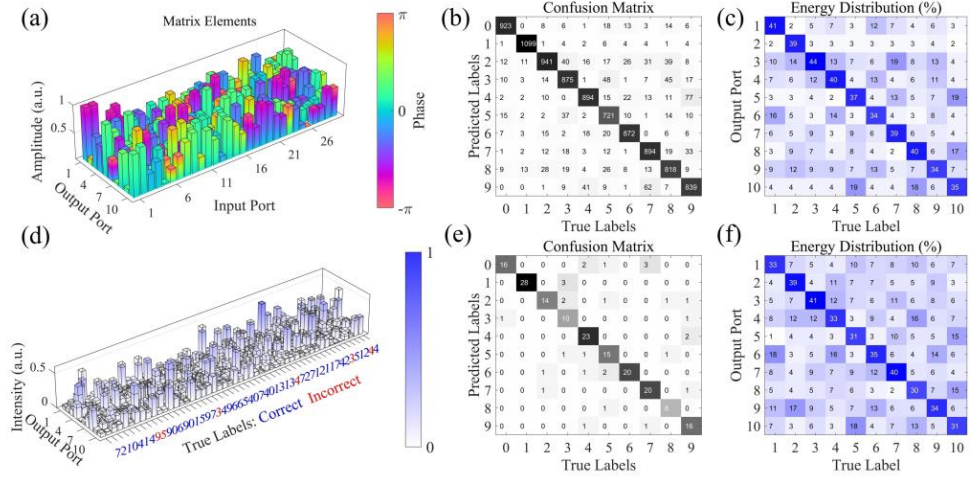

Fig.4. Results of MNIST digit task. (a) Complex amplitude of the 30 × 10 matrix. (b) and (c) Theoretical confusion matrix and energy distribution percentage over 10000 blind tests, respectively. (d) Measured intensity distribution of output neurons of 1~50 test images. (e) and (f) Measured confusion matrix and energy distribution percentage over 200 tests, respectively.

Similar experiments are performed on MNIST fashion task and the experimental results are summarized in Fig. 5. The complex amplitude of the optimized linear network weight is shown in Fig. 5(a), corresponding to a 37 × 10 complex matrix. Theoretical confusion matrix and energy distribution percentage over 10000 blind tests are shown in Fig. 5(b) and (c), respectively. Output intensity distribution of the 51~100 from 200 experiments are shown in Fig. 5(d), where filled bars and empty bars are corresponding to the measured and theoretical intensity distribution, respectively. The fidelity of this 37 × 10 network implementation is 0.952 ± 0.026 over 200 tests. 162 correct classification are observed, while the measured confusion matrix and energy distribution percentage are shown in Fig. 5 (e) and (f).

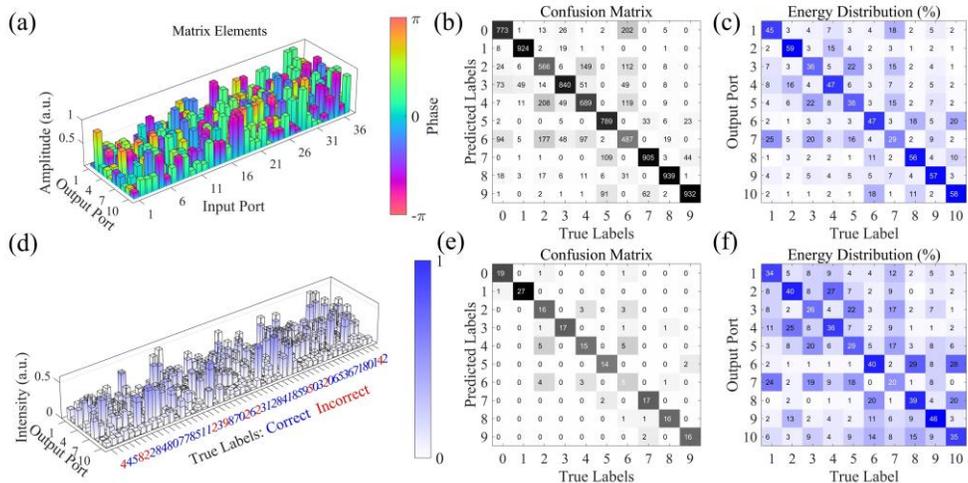

Fig.5. Results of MNIST fashion task. (a) Complex amplitude of the 37 × 10 matrix. (b) and (c) Theoretical confusion matrix and energy distribution percentage over 10000 blind tests, respectively. (d) Measured intensity distribution of output neurons of 51~100 test images. (e) and (f) Measured confusion matrix and energy distribution percentage over 200 tests, respectively.

The transformation weights of this neural network are complex numbers, and this is exactly the reason why square-law detection can perform the nonlinear activation. Moreover, the neural network is sensitive to phase information of the input image due to this feature. Besides MNIST tasks, another test has been made to examine the ability to recognize depth image [17] with our proposed architecture. The classification of topological charge of beam carrying orbital angular momentum (OAM) [18] is chosen as a concrete example. The spherical wave front of $\exp(il\varphi)$ is the representing characteristic of the $lth$ OAM mode, where $\varphi$ is the transverse angular coordinate. Notice that $\exp(il\varphi)$ is the phase term of a depth image and it is impossible to reveal phase information only from intensity distribution. To show this, OAM modes of topological charge $l$ ranging from 1 to 10 with uniform intensity distribution are generated and tested. For this task, the numbers of input neurons as well as output neurons are both ten. These neurons are aligned on a circle similar to those shown in Fig. 3(d). The transformation weights of this $10 \times 10$ network are actually the elements of ten-dimensional discrete Fourier transformation (DFT) matrix. It has been investigated in our previous work [19] that a DFT matrix of such configuration could serves as OAM mode recognizer. The measured output distribution of network for ten depth images of spherical phase is shown in Fig. 6(a). By choosing the label of the output port with maximum intensity, a 100% accuracy of OAM topological charge prediction is recorded. Three input OAM images with $l = 6,8,9$ as well as the corresponding output distribution are shown in Fig. 6(b). Although the input images have almost the same intensity and only differ in phase, our proposed all-optical neural network works quite well.

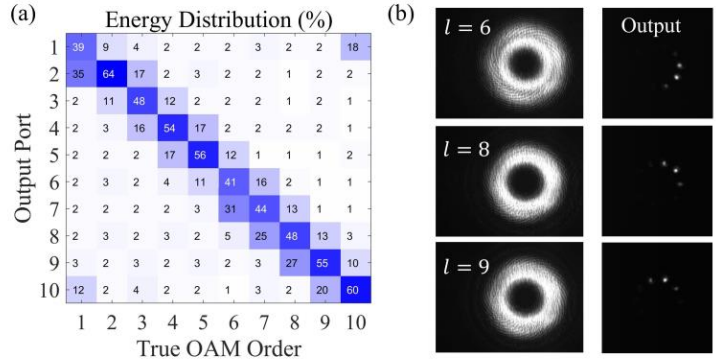

Fig.6. (a) Measured output distribution of network for ten depth images with spherical phase. (b) Three input OAM images with $l = 6,8,9$ and the corresponding output distribution

## 4. Discussion

Our work is some similar to the previously reported D²NN architecture [4], where only one nonlinear layer is sufficient for image classification tasks. According to our numerical calculation, the value of accuracy is a little bit lower than that of the D²NN. The main reason is the number of input neurons (or spatial sampling points) are much less in our work. However, this could be promoted by higher network dimensionality in future with larger SLM size and higher resolution. To achieve arbitrary network weights, the physical structure built by two SLMs (or phase masks) is independent to the matrix dimensionality in our design, which is the main difference to D²NN architecture. Besides, a simple training algorithm is employed in our work since the training process is not the main point of us. Our proposal focuses on implementing any pre-trained network weights with high fidelity in spite of the training algorithm.

According to experimental results, the fidelity is ~95%. The errors may come from misalignment of free space system, saturated effect of CCD camera, and the numerical estimation of spatial sampling effect. Two kinds of image sampling methods of 2D-interception and intensity sum are numerically compared with the configuration in Fig. 2 (c). The relative differences of optimized network weights are only around 1%, because the images in MNIST database do not contain much high spatial frequency components. Actually, the DCS modes are superpositions of individual Gaussian modes. A more precise model of sampling effect would include Gaussian convolution in future practical applications of more complicated image identification.

High processing speed is one the main feature of the ONN. In our proposal, the speed could be very fast due to the parallel nature of matrix multiplication. There are two factors to determine the speed in terms of the optical path delay and the photodetection rate. For the optical path delay, the propagation time of an input image in our architecture is ~2.7 ns, which is determined by the distance between SLM1 and the CCD camera (~0.8 m). Actually, this time could be further reduced. If the pipelining method is adopted, the minimum time interval between two images should only be longer than maximum optical path delay difference among different sampling points. In our presented setup, the maximum optical path difference of different DCS modes is ~0.125 mm, which is calculated from SLM distance (~0.4 m between SLM1 and SLM2) and the size of SLM active area ($8 \times 8\ mm^2$). This means that each image only needs to maintain for $4.17 \times 10^{-13}$ seconds, corresponding to a maximum switching speed of $S_o = 2.4 \times 10^{12}$ Hz. With this value, it could be found that the operation speed of all-optical matrix transformation could be very fast. However, considering the whole process of identification, input images have to be maintained before the photodetection is finished. Thus, the eventual speed of ONN is limited by the less one between the speed of the matrix transformation ($S_O$) and the photodetector ($S_D$). Similar to the equation in [2], we could estimate the number of options per second (floating point operations, FLOPs) required on a conventional computer to match the ONN proposed in this work:

$$R = 2 \times N \times M \times \min(S_O, S_D)\ \text{FLOPs} \qquad (4)$$

Here, $N \times M$ is the dimensionality of the transformation matrix while the factor of 2 is corresponding the complex number of each matrix element. Considering the dimensionality of implemented transformation matrix ($37 \times 10$) and the operation speed of CCD camera ($S_D = 100 Hz$), the processing speed of current setup can be estimated as only $R$=74K FLOPs. As mentioned and discussed above, the main limitation is the operation speed of detector. At state of art photodetector, photodetection rate could exceed 100GHz [20] and the corresponding $S_D = 10^{11} Hz$ is still much lower than the limitation induced by optical wave propagation $S_o = 2.4 \times 10^{12}$ Hz. Thus, our proposed ONN is potential to achieve much higher speed. Particularly, the processing speed could be as high as 7.4~74T FLOPs per second with 10~100GHz photodetector.

Furthermore, our proposed architecture could be scalable for deep neural network by employing all-optical nonlinear elements. Since the network weights are complex numbers, nonlinear effects such as electromagnetically induced transparency (EIT) [6] and nonlinear phase materials such as photorefractive crystal [5] are both available options. Moreover, an extension of all-optical training would also be possible with this architecture. The network weights are directly mapped to beam splitting ratios. There are simple and standard algorithms to generate the phase mask for particular beam splitting ratios as reported in our previous works [12,13] and others [21]. Thus, in optical training procedure, there is no necessary to refresh each pixel of SLMs one by one independently. Only the beam splitting ratios need to be refreshed, which is similar to the electrical training process, and then regenerate the phase masks. All-optical training may also solve the problem of imperfect implementation fidelity since the error corrections can be involved during optical training. The distance between phase masks could be decreased significantly with smaller pixel size. For a particular problem, the phase masks could also be fixed and prepared by metasurfaces [22] to obtain a compact and stable module. This is potential to be integrated with commercial cameras to enhance the image sensing abilities with low energy consumption.

**Funding.** National Key Research and Development Program of China (2017YFA0303700, 2018YFB2200402), the National Natural Science Foundation of China (Grant No. 61875101 and 61621064). This work was also supported by Beijing Innovation Center for Future Chip, Frontier Science Center for Quantum Information, Beijing academy of quantum information science, and Tsinghua University Initiative Scientific Research Program.

**Disclosures.** The authors declare no conflicts of interest.

**References**

1. B. J. Shastri, A. N. Tait, T. Ferreira de Lima, W. H. P. Pernice, H. Bhaskaran, C. D. Wright, and P. R. Prucnal, "Photonics for artificial intelligence and neuromorphic computing," Nat. Photonics **15**, 102–114 (2021).
2. Y. Shen, N. C. Harris, S. Skirlo, M. Prabhu, T. Baehr-Jones, M. Hochberg, X. Sun, S. Zhao, H. Larochelle, D. Englund, and M. Soljačić, "Deep learning with coherent nanophotonic circuits," Nature Photon **11**, 441–446 (2017).
3. A. N. Tait, M. A. Nahmias, B. J. Shastri, and P. R. Prucnal, "Broadcast and Weight: An Integrated Network For Scalable Photonic Spike Processing," J. Lightwave Technol. **32**, 4029–4041 (2014).
4. X. Lin, Y. Rivenson, N. T. Yardimci, M. Veli, Y. Luo, M. Jarrahi, and A. Ozcan, "All-optical machine learning using diffractive deep neural networks," Science **361**, 1004–1008 (2018).
5. T. Yan, J. Wu, T. Zhou, H. Xie, F. Xu, J. Fan, L. Fang, X. Lin, and Q. Dai, "Fourier-space Diffractive Deep Neural Network," Phys. Rev. Lett. **123**, 023901 (2019).
6. Y. Zuo, B. Li, Y. Zhao, Y. Jiang, Y.-C. Chen, P. Chen, G.-B. Jo, J. Liu, and S. Du, "All-optical neural network with nonlinear activation functions," 6 (n.d.).
7. J. Feldmann, N. Youngblood, C. D. Wright, H. Bhaskaran, and W. H. P. Pernice, "All-optical spiking neurosynaptic networks with self-learning capabilities," Nature **569**, 208–214 (2019).
8. X. Xu, M. Tan, B. Corcoran, J. Wu, A. Boes, T. G. Nguyen, S. T. Chu, B. E. Little, D. G. Hicks, R. Morandotti, A. Mitchell, and D. J. Moss, "11 TOPS photonic convolutional accelerator for optical neural networks," Nature **589**, 44–51 (2021).
9. M. A. Nahmias, B. J. Shastri, A. N. Tait, and P. R. Prucnal, "A Leaky Integrate-and-Fire Laser Neuron for Ultrafast Cognitive Computing," IEEE J. Select. Topics Quantum Electron. **19**, 1–12 (2013).
10. I. A. D. Williamson, T. W. Hughes, M. Minkov, B. Bartlett, S. Pai, and S. Fan, "Reprogrammable Electro-Optic Nonlinear Activation Functions for Optical Neural Networks," IEEE J. Select. Topics Quantum Electron. **26**, 1–12 (2020).
11. M. T. Hill, E. E. E. Frietman, H. de Waardt, Giok-djan Khoe, and H. J. S. Dorren, "All fiber-optic neural network using coupled SOA based ring lasers," IEEE Trans. Neural Netw. **13**, 1504–1513 (2002).
12. S. Li, S. Zhang, X. Feng, S. M. Barnett, W. Zhang, K. Cui, F. Liu, and Y. Huang, "Programmable Coherent Linear Quantum Operations with High-Dimensional Optical Spatial Modes," Phys. Rev. Applied **14**, 024027 (2020).
13. P. Zhao, S. Li, X. Feng, S. M. Barnett, W. Zhang, K. Cui, F. Liu, and Y. Huang, "Universal linear optical operations on discrete phase-coherent spatial modes with a fixed and non-cascaded setup," J. Opt. **21**, 104003 (2019).
14. W. R. Clements, P. C. Humphreys, B. J. Metcalf, W. S. Kolthammer, and I. A. Walmsley, "Optimal design for universal multiport interferometers," Optica **3**, 1460 (2016).
15. Y. LeCun, Y. Bengio, and G. Hinton, "Deep learning," Nature **521**, 436–444 (2015).
16. M. A. Nielsen and I. L. Chuang, *Quantum Computation and Quantum Information*, 10th anniversary ed (Cambridge University Press, 2010).
17. W. Aly, S. Aly, and S. Almotairi, "User-Independent American Sign Language Alphabet Recognition Based on Depth Image and PCANet Features," IEEE Access **7**, 123138–123150 (2019).
18. L. Allen, M. W. Beijersbergen, R. J. C. Spreeuw, and J. P. Woerdman, "Orbital angular momentum of light and the transformation of Laguerre-Gaussian laser modes," Physical Review A **45**, 8185–8189 (1992).
19. Y. Wang, V. Potoček, S. M. Barnett, and X. Feng, "Programmable holographic technique for implementing unitary and nonunitary transformations," Phys. Rev. A **95**, 033827 (2017).
20. L. Vivien, A. Polzer, D. Marris-Morini, J. Osmond, J. M. Hartmann, P. Crozat, E. Cassan, C. Kopp, H. Zimmermann, and J. M. Fédéli, "Zero-bias 40Gbit/s germanium waveguide photodetector on silicon," 6 (2012).
21. J. Müller-Quade, H. Aagedal, Th. Beth, and M. Schmid, "Algorithmic design of diffractive optical systems for information processing," Physica D: Nonlinear Phenomena **120**, 196–205 (1998).
22. G. Li, S. Zhang, and T. Zentgraf, "Nonlinear photonic metasurfaces," Nat Rev Mater **2**, 17010 (2017).